\documentstyle[preprint,aps]{revtex}

\begin{document}
%\today
%\tighten
\draft
\preprint{SNUTP 93-68}
\title{Fermion Doubling and a Natural Solution
of  the Strong CP Problem}
\author{Sanghyeon Chang \ and \ Jihn E. Kim}
\address{Department of Physics and Center for Theoretical Physics\\
 Seoul National University, Seoul 151-742, Korea}

\maketitle
\begin{abstract}
We suggest the fermion doubling for all quarks and leptons.
It is a generalization of
the neutrino doubling of the seesaw mechanism.  The new quarks
and leptons are $SU(2)$ singlets and carry the electromagnetic
charges of their lighter counterparts.  An $SU(3)$ {\it anomaly free
global symmetry} or a discrete symmetry
can be introduced to restrict the Yukawa couplings.
The form of mass matrix is belonging to that of
Nelson and Barr even though our model does not belong to Barr's criterion.
The weak CP violation of the Kobayashi-Maskawa form
is obtained through the spontaneous breaking of CP symmetry at high
energy scale.  The strong CP solution is through a specific form
of the mass matrix.
At low energy, the particle content is the same as in the standard model.
For a model with a global symmetry, in addition there exists
a massless majoron.

\end{abstract}

\narrowtext
\section{Introduction}
The standard model is remarkably successful in describing the
electroweak phenomena.  However, it has twenty or so unexplained
parameters.  Two of these parameters, $\bar{\theta}=\theta_{QCD}
+\theta_{QFD}$ and the weak
CP phase, are related to the CP symmetry.  The strong
CP problem has several candidate solutions: axions \cite{pq,axion1,axion2}
 and natural solutions \cite{georgi,nelson,bar,bbp}. The weak CP phase will
be understood if the electroweak symmetry breaking is unveiled.
At present, there are several
candidates for the realization of weak CP violation, the
Kobayashi-Maskawa model, superweak CP violation, Higgs mediated
CP violation, etc.  These can be classified into explicit CP
violation (including hard CP violation if complex Yukawa couplings
are introduced and soft CP violation if complex scalar masses \cite{georgi}
are introduced) and spontaneous CP violation.

The most plausible model among natural solutions is one which was proposed
by Nelson \cite{nelson} and generalized by Barr \cite{bar}.
This kind of models assumes a CP invariant Lagrangian, and the weak CP
violation
is introduced by spontaneous breaking of CP symmetry.
A key point of the Nelson--Barr type models is an introduction of a heavy scale
(heavy scalar and heavy fermion). Nelson assumes a specific form of mass matrix
so that its determinant remains real after the spontaneous CP violation.
If we have $N$ heavy quarks $Q_i$ and 3 light generations,
the Nelson--Barr form of the fermion mass matrix at tree level is
\begin{equation}
\left(\begin{array}{cc} hv&0\\ \alpha&M\end{array}\right)
\end{equation}
where $hv$ is a $3\times 3$, $M$ is a $N\times N$ real matrix and $\alpha$ is
a $N\times 3$ complex matrix, in the broken phase.
Determinant of this mass matrix is real, even though the matrix, i.~e. the
$\alpha$ entry, is complex.
However, after the spontaneous CP breaking, there is no symmetry that keeps
the determinant of the mass matrix real. So a small $\bar{\theta}$ can be
generated from radiative corrections.
Since $|\bar{\theta}|$ is less than $10^{-9}$, smallness of  the gauge and/or
Yukawa couplings is required. Thus it seems that the Nelson--Barr model
reintroduces the hierarchy problem, i.~e. the smallness of parameters.
To alleviate the restriction on coupling constants, the supersymmetrization of
the Nelson--Barr model was suggested \cite{seger}. However, the
supersymmetrization does not improve this difficulty, due to an entry of the
gluino
mass \cite{dine}. Therefore, at present the smallnesses of some couplings are
required in both nonsupersymmetric and supersymmetric Nelson--Barr type models.

Here, we do not attempt to solve this fundamental problem, but unify
the strong CP solution {\it \`{a} la} Nelson and Barr and the seesaw mechanism.
The seesaw model is the most popular and plausible mechanism of the small
neutrino mass generation. The seesaw mechanism also employs two
scales, the heavy neutrino scale and the weak scale \cite{seesaw}.
To give a non--zero mass to each neutrino, a heavy right--handed singlet
neutrino must be introduced for each $SU(2)_L$ lepton doublet.
These heavy neutrinos have no logical relation with the heavy quarks in the
Nelson--Barr type model, because the number of the heavy fermions does not
depends on the number of light fermions and the couplings
between the $SU(2)_L$ doublets and heavy right--handed singlet fermions are
forbidden in the Nelson--Barr type model.

In this paper, we suggest {\it fermion doubling}, which can relate the
heavy scale in the Nelson--Barr model to the heavy neutrino scale in
the seesaw model.  There is no fundamental theory which explains doubling of
fermions.
However, the fermion doubling gives us a consistent view of the unified
solution
of these two different problems. If we invent a mechanism which has two scales,
a heavy fermion partner at a heavy scale for each fermion at the weak scale,
it would explain the reasons `why the $\bar{\theta}$ is so small' and `why the
neutrino mass is so small'. The fundamental theory may contain one solution to
various mysteries of low energy phenomena. Namely, it is worth to make a
phenomenologically viable model that unifies phenomenologically unrelated
problems. \footnote{
A similar idea was realized within the scheme of the left--right symmetric
model
of Babu and Mohapatra \cite{moha}.}

For the desired mass matrix to appear, there must exist a symmetry. In Sec. II,
we achieve this with a global symmetry.
In our previous paper \cite{chang}, we suggested a similar model with an extra
heavy quark
$Q$ in addition to the fermion doubling with a global symmetry so that a heavy
quark axion is introduced for a solution of the strong CP problem.
This model was suggested to unify the heavy quark axion scale and the neutrino
seesaw scale.
In that case, $U(1)_{PQ}$ symmetry which is broken at majoron scale
is necessarily anomalous. That was the reason why we introduced
the additional  heavy quark $Q$.  Without the heavy quark $Q$,
the $U(1)$ symmetry would not be the Peccei-Quinn symmetry, and the resulting
Goldstone boson would be a practically massless majoron.  At low energy
we have no particle other than those in the contents of minimal
standard model except the majoron, and the low energy phenomenology
is the same as the standard model.

In Sec. III, we introduce a $Z_2$
symmetry instead of the global symmetry. Sec. IV is a conclusion.
There are two appendixes. In Appendix A, we reviewed the
generation of Kobayashi--Maskawa CP phase from the mass matrix at the low
energy. In appendix B, we present one loop calculation of
$\bar{\theta}$.

\section{A Model with a global symmetry}
By the fermion
doubling, we introduce a new $SU(2)$ singlet fermion for each fermion
of the standard model.  We also introduce two $SU(2)$ singlet
complex scalars, $\sigma$, and $S$, to realize a desirable
intermediate scale physics.
Then fermion contents of the model become,
\begin{eqnarray}
\mbox{The quark sector }&:&q^i_L,u^i_R,d^i_R, U^i_L,U^i_R,D^i_L,D^i_R,
\nonumber\\
\mbox{The lepton sector}&:&l^i_L,e^i_R,N^i_R,E^i_L,E^i_R,
\end{eqnarray}
where $q_L$ and $l_L$ are $SU(2)$ doublets and $i(=1,2,3)$ is the family
index. The particles added are denoted as capital letters. Their
electromagnetic
charges are the same as those of the corresponding light
particles (lower case symbols).  We assign $U(1)_{global}$ charges to
the fermions as,
\begin{eqnarray}
1&\mbox{ for }&  U_R, D_R, E_R, N_R, q_L, l_L,\nonumber\\
-1&\mbox{ for }&  U_L, D_L, E_L, u_R, d_R, e_R,\nonumber\\
0&\mbox{ for }& H,\nonumber\\
-2&\mbox{ for }& \sigma, S .
\end{eqnarray}
Note that the Higgs doublet $H$ carries a vanishing global charge.
The global symmetry does not have an $SU(3)$ anomaly and hence
cannot be the Peccei--Quinn symmetry. \footnote{
But, there exists the $U(1)_{global}-SU(2)-SU(2)$ anomaly.  If the
unknown quantum gravity effects spoil the
needed global symmetry badly, one has to remove this anomaly.
But at present, it is not serious because we
are not sure how the quantum gravity dictates the low energy
phenomenology. However, it is easy to remove the $U(1)_{\rm global}
-SU(2)-SU(2)$ anomaly by simply introducing six $SU(2)$ triplets
$L_A\equiv(L^+_A,L^0_A,L^-_A)_L,$ $(A=1,\cdots,6)$ and assigning $-1$
for the $U(1)_{\rm global}$ charges. The couplings of the form
$(b_{L_A}\sigma^*+c_{L_A}S^*)L_A^TC^{-1}L_A$ give the intermediate scale masses
to $L_A$'s where $C\equiv \exp(i\pi T_2)$.
In addition, if one tries to make Tr $Y_{global}=0$, where $Y_{global}$ is the
generator of $U(1)_{global}$, he can introduce additional gauge singlets
with non-vanishing $U(1)_{global}$ charge.
The hypothetical $L$ particles are not stable due to the coupling of the form
$({\tilde H}^T l)_{sym} L$.}

The renormalizable Lagrangian obeying the $SU(3)\times SU(2)\times U(1)_Y
\times U(1)_{global}\times CP$ symmetry can be written as
\widetext
\begin{eqnarray}
{\cal L}&=& (b_U\sigma+c_US)
\bar{U}_LU_R+(b_D\sigma+c_DS)\bar{D}_LD_R+
(b_E\sigma+c_ES) \bar{E}_LE_R+
\frac{1}{2}(b_N\sigma+c_NS)\bar{N}^c_R N_R\nonumber\\
&&+h_U^T\bar{U}_R\tilde{H}^Tq_L +h_D^T\bar{D}_RH^Tq_L
+h_N^T\bar{N}_R\tilde{H}^Tl_L +h_E^T\bar{E}_RH^Tl_L
\nonumber\\&&
+\alpha_U \bar{U}_Lu_R+\alpha_D\bar{D}_Ld_R
+\alpha_E\bar{E}_Le_R+{\rm h. c.}\nonumber\\
&&-V(\sigma,S,H)+{\cal L}_{kinetic}+{\cal L}_{gauge}
\label{1},\end{eqnarray}
\narrowtext
\noindent where $H$ is a Higgs doublet scalar with $Y=-1/2$,
 ${\tilde H}=i\sigma_2 H^*$. In Eq.~(\ref{1}), we suppressed flavor indices.
For example, $h_U^T\bar{U}_R\tilde{H}^Tq_L$ means
$(h_U)_{ji}\bar{U_i}_R\tilde{H}^T{q_j}_L=\bar{U}_R\tilde{H}^T
h^T_Uq_L$.
All the coupling matrices are real due to the CP symmetry and
$\theta_{QCD}=0$.  The potential is
\widetext
\begin{eqnarray}
V(\sigma,S,H)&=&\mu^2_HH^\dagger H+\mu^2_\sigma\sigma^*\sigma
+\mu^2_SS^*S+\mu^2_{\sigma S}(\sigma^* S+ S^*\sigma)+
\lambda_H(H^\dagger H)^2
+\lambda_\sigma(\sigma^*\sigma)^2
\nonumber\\ &&
+ \lambda_S(S^*S)^2
+\lambda_{\sigma S}((S^*\sigma)^2+(\sigma^* S)^2)
+\tilde{\lambda}_{\sigma S}|\sigma^* S|^2
+\lambda_{\sigma SS}(S^*S)(\sigma^* S+S^*\sigma)
\nonumber\\ &&
+\lambda_{\sigma\sigma S}(\sigma^*\sigma)(\sigma^* S+S^*\sigma)
+ \lambda_{\sigma H}H^\dagger H\sigma^*\sigma
+\lambda_2H^\dagger HS^*S
\nonumber\\ &&
+\lambda_{\sigma 2}H^\dagger H(\sigma^*S+S^*\sigma)
.\end{eqnarray}
\narrowtext
\noindent where the range of parameters are chosen so that the following
vacuum expectation values develop at the symmetry breaking minimum,
\begin{eqnarray}
\sigma&=&\frac{\tilde{v}_1+\tilde\rho_1}{\sqrt{2}}e^{ia_1/\tilde{v}_1},\ \
S=\frac{\tilde{v}_2+\tilde\rho_2}{\sqrt{2}}e^{ia_2/\tilde{v}_2},\nonumber\\
H&=&\frac{1}{\sqrt{2}}\left(\begin{array}{c}0\\v+\rho\end{array}\right)
e^{i\phi/v}.
\end{eqnarray}
Here $\tilde{v}_1,\tilde{v}_2$ are the seesaw scale (intermediate scale)
and $v$ is the weak scale.  Note that $\alpha_F$ can be of order
$\tilde v$ where $\tilde v \equiv \sqrt{|\tilde{v}_1|^2+|\tilde{v}_2|^2}$.

{\it The weak CP violation occurs through spontaneous
symmetry breaking.}  Because of the $\mu_{\sigma S}^2$, $\lambda_{\sigma
SS}$ and $\lambda_{\sigma SS}$ terms,
$<\sigma>$ and $<S>$ can be complex, and weak CP violation is
generated.  The essential points of this type of models are that
diagonalization of {\it the fermion mass matrix
introduces the Kobayashi--Maskawa CP phase which is not suppressed by
$v/ \tilde v$ and vanishing} Arg Det $M_q$. (See Appendix.)

The tree level mass matrix of fermions except for neutrinos written
in the \\ $(f^1,f^2,f^3,F^1,F^2,F^3)_L$ $\otimes
(f^1,f^2,f^3,F^1,F^2,F^3)_R$ space is

\begin{equation}
M_f=\left(\begin{array}{cc}0&h_Fv\\ \alpha_F&b_F\tilde{v}_1+
c_F\tilde{v}_2\end{array}\right),
\end{equation}
where $b,\ c,\ h,\ \alpha$ are $3\times3$ matrices and $f$
and $F$ represent approximately light and
heavy fermions, respectively.
{}From now on, we set
$\beta\equiv hv$ and $\Omega\equiv b_F\tilde{v}_1+c_F\tilde{v}_2$.
Here, we cannot remove the phase between $\tilde{v}_1$ and
$\tilde{v}_2$. So $\Omega$ is a general $3\times 3$ complex
matrix, and $\alpha$ and $\beta$ are general
$3\times 3$ real matrices.
Then the quark mass matrix is

\begin{equation}
M_q\equiv\left(
\begin{array}{cc}
0&\beta\\
\alpha&\Omega
\end{array}
\right) \label{mass}
\end{equation}
where $\beta$ is of order electroweak scale and $\alpha$ and $\Omega$
are at the intermediate mass scale, presumably at $10^{10}-10^{13}$
GeV.
We assume $O(\alpha)\lesssim O(\Omega)$.

The quark mass matrix $M_q$ of Eq.~(\ref{mass}) gives
\begin{equation}
\mbox{Arg Det } M_q = 0 .
\end{equation}
Note that complex phases appear only in the $\Omega$ block.  However,
they do not contribute to Arg Det $M$ due to 0 entries at the upper
left corner.  Thus even after the introduction of the weak CP phase,
$\theta_{QFD}$ remains zero.

Even though $\theta_{QFD}$ is zero at tree level, the radiative
corrections to the quark mass matrix can be complex in general.
The complex phases
occur at heavy quark sectors only through the spontaneous CP
violation, but after the symmetry breaking, the dimension 5 operator
arising at loop order, $(1/M)S(g_u\bar{u}\tilde{H}^T+g_d\bar{d}H^T)q_L $,
may give a complex correction to the light--light block of the fermion
mass matrices (see the $0$ entry in Eq.~(\ref{mass})) and thus
to the Arg Det $M_q$. However, the estimation is not so trivial.
If there is no other boson, i.~e.~other than $H$, which mixes
the light fermions and heavy fermions, the one loop corrections are
generated only by the Feynman diagram in Fig. 1.
This diagram always contains
\begin{equation}
\Gamma M^{-1}MP_l\Psi_R,
\end{equation}
term, where
\begin{eqnarray}
\Gamma=\left(\begin{array}{cc} 0&0\\ 0&b_F\end{array}\right)
\mbox{, }\left(\begin{array}{cc} 0&0\\ 0&c_F\end{array}\right)
\mbox{ or }\left(\begin{array}{cc} 0&h_F\\ 0&0\end{array}\right),
\nonumber\\
M=\frac{(1-\gamma_5)}{2}M_q+\frac{(1+\gamma_5)}{2}M_q^\dagger, \nonumber\\
\mbox{and\ \ } P_l=\left(\begin{array}{cc} 1&0\\ 0&0\end{array}\right)
,\ \Psi_R=\left(\begin{array}{c} f_R\\ F_R\end{array}\right).\label{LR}
\end{eqnarray}
Since this term always vanishes trivially,
there is no one loop contribution to the strong CP phase
with our particle contents.
One loop correction to $\bar{\theta}$ comes from either one loop diagram
with heavy GUT gauge bosons which mix the light and the heavy fermion families,
as the original Nelson--Barr model,
or $U(1)_{global}$ neutral singlet scalar which couples to $\bar{F}_Lf_R$.

For the neutrinos, the mass matrix can
be written in $(\nu_L, N_R)\otimes(\nu_L, N_R)$
basis as,
\begin{equation}
M_N=\left(\begin{array}{cc}0&h_Nv\\ h^T_Nv&b_N\tilde{v}_1+c_N\tilde{v}_2
\end{array}\right).
\end{equation}
Diagonalizing this mass matrix, one can see that a light neutrino acquires  the
mass
$m_\nu\simeq (h_Nv)^2/\tilde{v}$ which is very small.  Thus the neutrino
mass has a further suppression.  This is because for the quark and
charged lepton sectors we have the $\alpha$ matrix of order $\tilde v$
while for the neutrino sector the corresponding block is of order $v$.
Thus light neutrinos have mass at order $v\times (v/\tilde v)$.
At low energy, there exists a massless Goldstone boson, majoron, in addition
to the standard model particles because the global $U(1)$ symmetry is
broken spontaneously at the heavy neutrino scale.

\section{A $Z_2$ Model}
The minimal Nelson--Barr type model was studied by Bento et.~al.~\cite{bbp}.
Here, we can use a discrete symmetry to get the same mass matrix as
Eq.~(\ref{mass}).
The simplest example is the model with a $Z_2$ symmetry.
In this case, we can construct the model with one complex scalar singlet.
We assign $Z_2$ charges to the fermions as,
\begin{eqnarray}
0&\mbox{ for }&  U_R, D_R, E_R, N_R, q_L, l_L,\nonumber\\
1&\mbox{ for }&  U_L, D_L, E_L, u_R, d_R, e_R,\nonumber\\
0&\mbox{ for }& H,\nonumber\\
1&\mbox{ for }& S .
\end{eqnarray}

The renormalizable Lagrangian obeying the $SU(3)\times SU(2)\times U(1)_Y
\times Z_2\times CP$ symmetry can be written as
\widetext
\begin{eqnarray}
{\cal L}&=& (b_US+c_US^*)
\bar{U}_LU_R+(b_DS+c_DS^*)\bar{D}_LD_R+
(b_ES+c_ES^*) \bar{E}_LE_R+
\frac{1}{2}M_\nu\bar{N}^c_R N_R\nonumber\\
&&+h_U^T\bar{U}_R\tilde{H}^Tq_L +h_D^T\bar{D}_RH^Tq_L
+h_N^T\bar{N}_R\tilde{H}^Tl_L +h_E^T\bar{E}_RH^Tl_L
\nonumber\\&&
+\alpha_U \bar{U}_Lu_R+\alpha_D\bar{D}_Ld_R
+\alpha_E\bar{E}_Le_R+{\rm h. c.}\nonumber\\
&&-V(S,H)+{\cal L}_{kinetic}+{\cal L}_{gauge}
\label{2},\end{eqnarray}
\narrowtext
All the coupling matrices are real due to the CP symmetry and
$\theta_{QCD}=0$. One can see that lepton number is not a good symmetry of
the Lagrangian. Thus, there is no Goldstone boson like majoron. The potential
is
\widetext
\begin{eqnarray}
V(S,H)&=&\mu^2_HH^\dagger H
+\mu^2_SS^*S+ \lambda_H(H^\dagger H)^2
+ \lambda_S(S^*S)^2
+\lambda_1H^\dagger HS^*S
\nonumber\\ &&
+\lambda_2H^\dagger H(S^2+S^{*2})
+\lambda_3(S^2+S^{*2})+\lambda_4SS^*(S^2+S^{*2})+\lambda_5(S^4+S^{*4})
.\label{pot}\end{eqnarray}
\narrowtext
\noindent where the range of parameters are chosen so that the following
vacuum expectation values develop at the symmetry breaking minimum,
\begin{equation}
S=\frac{\tilde{v}+\tilde\rho}{\sqrt{2}}e^{ia/\tilde{v}},\
H=\frac{1}{\sqrt{2}}\left(\begin{array}{c}0\\v+\rho\end{array}\right)
e^{i\phi/v}.
\end{equation}
Here $\tilde{v}$ and $M_\nu$ are the intermediate scales
and $v$ is the weak scale.

 The weak CP violation occurs through spontaneous
symmetry breaking, as in the last section.
Here, $<S>$ can be complex without introducing further scalar singlets
and the weak CP violation is generated spontaneously.

The tree level mass matrix of quarks can be written
in the \\ $(f^1,f^2,f^3,F^1,F^2,F^3)_L$ $\otimes
(f^1,f^2,f^3,F^1,F^2,F^3)_R$ space as

\begin{equation}
M_f=\left(\begin{array}{cc}0&h_Fv\\ \alpha_F&b_F\tilde{v}+
c_F\tilde{v}^*\end{array}\right).
\end{equation}
We set
$\beta\equiv hv$ and $\Omega\equiv b_F\tilde{v}+c_F\tilde{v}^*$.
 Then $\Omega$ is a general $3\times 3$ complex
matrix, and $\alpha$ and $\beta$ are general
$3\times 3$ real matrices.
Then the quark mass matrix is

\begin{equation}
M_q\equiv\left(
\begin{array}{cc}
0&\beta\\
\alpha&\Omega
\end{array}
\right)
\end{equation}
where $\beta$ is of order electroweak scale and $\alpha$ and $\Omega$
are at the intermediate mass scale, $10^{10}-10^{13}$ GeV.

In $Z_2$ symmetry model, non-vanishing CP phases come from neutral Higgs
one loop corrections to
the quark self energy diagrams (Fig. 2), which will give complex contributions
to the light--heavy block in the mass matrix. We have four such diagrams.
{}From Fig. 2, adding all the one loop contributions, we have
\begin{equation}
g(\lambda_1bc(S^2+S^{*2})+\lambda_2(b^2 S^{*2}+ c^2 S^2))
\tilde{D_R}H^Tq_L,
\end{equation}
where $g\sim\frac{1}{16\pi^2\tilde{v}^2}$.
 (The family indices $U,D$ of the coupling matrices $b$ and $c$ are omitted.)
After symmetry breaking, $\theta_{QFD}$ can be estimated as,
\begin{equation}
\theta_{QFD}\sim\frac{1}{16\pi^2}\lambda_2(\tilde{b}^2-\tilde{c}^2),
\end{equation}
where $\tilde{b}$ and $\tilde{c}$ are leading order values of the
matrices $b$ and $c$.
This shows that the strong CP phase $\bar \theta$ depends on the
coupling constants of the $H-S$ terms and the Yukawa couplings.
At the weak scale  $v$,
the bare mass of the Higgs doublet is
\begin{equation}
\mu^2=\mu^2_H+\lambda_1 |\tilde{v}|^2+\lambda_22Re\tilde{v}^2.
\end{equation}
If  $\lambda$'s are not small we need huge cancellation in Eq.~(\ref{pot}).
 So, it is not unnatural to think $\lambda_2\sim v^2/\tilde{v}^2$.
In this case, $|\theta_{QFD}|$ is less than $10^{-20}$.
$\lambda_2$ as large as $10^{-7}$ would not contradict the requirement
$|\bar{\theta}|<10^{-9}$.
The requirement of small $\lambda_2$ belongs to the category of the gauge
hierarchy problem which was pointed out by Bento et.~al. \cite{bbp}.
One may consider this requirement is different from the requirement of
a small $\bar \theta$.

The neutrino mass matrix can
be written in $(\nu_L, N_R)\otimes(\nu_L, N_R)$
basis as,
\begin{equation}
M_N=\left(\begin{array}{cc}0&h_Nv\\ h^T_Nv&M_\nu
\end{array}\right).
\end{equation}
The diagonalized light neutrino mass is
$m_\nu\simeq (h_Nv)^2/M_\nu$.

\section{Conclusion}
In conclusion, a simple extension of the seesaw mechanism to the quark
sector by fermion doubling can give a natural solution of the strong CP problem
by the same argument as that of Nelson and Barr.
We introduced a heavy fermion for every light fermion. To have a desired
fermion mass matrix, a symmetry, a $U(1)_{global}$ or a discrete symmetry
was introduced.  CP is a symmetry
of the Lagrangian and the weak CP violation is introduced by
spontaneous symmetry breaking.   At low energy, the weak CP
phenomenology is the same as the Kobayashi--Maskawa model, which is  the most
attractive feature of the Nelson--Barr type models among natural solutions.
The correction to the strong CP phase comes from at one loop order
in the fermion mass matrix. If, however, we do not consider the
heavy gauge bosons which will mix light and heavy quarks, there is
no one loop correction term in the model with the additional
global symmetry. But if we consider discrete symmetry or GUT gauge
bosons, the strong CP phase appears at one loop order.
 Our model for the solution
of the strong CP problem belongs to the category of natural solutions.
It does not belong to the mass matrix of the Nelson--Barr ansatz,
since it admits
couplings between the light $SU(2)_L$ doublets and heavy singlet fermions,
 but its spirit is the same as theirs. The way of suppressing strong CP
phase and generating KM phase is the same as that of Nelson and Barr.
{\it This unique feature comes from the fact
that the numbers of heavy fermions and light fermions are the
same in our case}. Therefore we can couple light and heavy family with
Higg's doublet and forbid the light fermion--light fermion couplings,
but still obtain a phenomenologically successful fermion mass matrix.
This modification does not break the merits of the Nelson--Barr type
model. In addition, we observe that if the number of heavy fermions is the
same as that of light fermions in the Nelson--Barr model,
one can interchange $3\times 3$ blocks in $6\times 6$ quark mass matrix,
\begin{equation}
\left(\begin{array}{cc}\beta&0\\ \alpha&\Omega\end{array}\right)
\longrightarrow
\left(\begin{array}{cc}0&\beta\\ \Omega&\alpha\end{array}\right).
\end{equation}
In this form, it is equivalent to our model, and does not affect the tree
level properties of the Nelson--Barr type model.
However, it is not so obvious to tell whether one loop
correction of our model is not different from that of Nelson--Barr type model.
In Appendix B, we show that the one loop calculations
of $\bar{\theta}$ in both models have the same magnitude at leading order.

We showed two examples in the fermion doubling scheme, one with global $U(1)$
and the other with $Z_2$.
In either case, fermion doubling provides a unified way which has a
consistent and simple explanation of the small
strong CP phase and the small neutrino mass generation by the seesaw mechanism.

\acknowledgments
This work was supported in part by the Korea Science and Engineering
Foundation through Center for Theoretical Physics,
Seoul National University, KOSEF--DFG Collaboration Program,
and the Ministry of Education.

\appendix
\section{generation of kobayashi--maskawa CP phase}
In this appendix, we show that the unsuppressed Kobayashi--Maskawa CP phase can
be generated from the fermion mass matrix of the fermion doubling model.
Actually the square matrix $M_qM_q^\dagger$ is exactly same as that of the
Nelson--Barr
model. In this sense, this appendix is a review of the Nelson--Barr type model.
We can diagonalize this $6\times 6$ matrix by biunitary transformation,
\begin{equation}
U_L M_q U^\dagger_R=M_q^D.
\end{equation}
where $M_q^D$ is a diagonal matrix.
Next step is to diagonalize the square of the up quark
matrix $M_u M_u^\dagger$ by a unitary matrix $U^u_L$ and down quark matrix
$M_d^\dagger M_d$ by $U_R^d$.
Then the light quark mixing matrix $K$ is
the upper left corner $3\times 3$ sub-matrix  of $U^u_L
{U^d_R}^\dagger$.  As shown below, this $K$ is not a
unitary matrix, but it becomes unitary in the limit of
$v/\tilde v=0$.
  More importantly, in the limit of $v/\tilde v=0$ this submatrix $K$ is not
real orthogonal but unitary.
 Let us express $U_L^u$ and $M_q^D$ in terms of $3\times 3$ matrices $C_1$,
$S_1$,
 etc,
\begin{equation}
U^u_L=\left(
\begin{array}{cc}
C_1&S_1\\
S_2&C_2
\end{array}
\right),\
M_q^D=\left(
\begin{array}{cc}
m_q&0\\
0&m_Q
\end{array}
\right),
\end{equation}
where $m_q$ and $m_Q$ are the diagonal light and heavy quark matrices,
respectively.  A Hermitian matrix $H$ ($M_q^\dagger M_q$ or $M_q M_q^\dagger$)
can be diagonalized by a unitary transformation $UHU^\dagger$.
To show that $K$ is complex, we will prove below that $C_1$
is unitary in the limit of $v/\tilde{v}=0$.
Diagonalization of the matrix $M_q M_q^\dagger$ gives the
following conditions,
\widetext
\begin{mathletters}
\begin{equation}
C_1\beta\beta^T C_1^\dagger  + C_1\beta\Omega^\dagger S_1^\dagger  +
S_1\Omega\beta^T C_1^\dagger  +S_1(\Omega\Omega^\dagger+
\alpha\alpha^T)S_1^\dagger =m_qm_q^\dagger,\label{m}
\end{equation}
\begin{equation}
S_2\beta\beta^T S_2^\dagger  + S_2\beta\Omega^\dagger C_2^\dagger  +
C_2\Omega\beta^T S_2^\dagger  +C_2(\Omega\Omega^\dagger+
\alpha\alpha^T)C_2^\dagger =m_Qm_Q^\dagger,\label{M}
\end{equation}
\begin{equation}
C_1\beta\beta^T S_2^\dagger + C_1\beta\Omega^\dagger C_2^\dagger+
S_1\Omega\beta^T S_2^\dagger + S_1(\Omega\Omega^\dagger+
\alpha\alpha^T)C_2^\dagger=0.\label{zero}
\end{equation}
\end{mathletters}
\narrowtext
Pay attention to Eq.~(\ref{m}).  If $v/\tilde{v}$ is very small,
the components of $S_1$ are at most of order $v/\tilde v$,
since physical quarks masses are smaller than the electroweak scale.
It can be easily seen that the $3\times 3$ matrices $C_{1,2}$
become some unitary matrices and $S_2$ are of order $v/\tilde{v}$
as $v/\tilde{v}$ goes to zero,
due to the unitarity condition of the $6\times 6$ matrix $U_L^u$.
{}From Eq.~(\ref{zero}) we find,
\begin{equation}
O(S_{1,2}\Omega)\simeq O(C_{1,2}\beta)\simeq O(\beta).
\end{equation}
It is readily noticed that all four components of
Eq.~(\ref{m}) do contribute even in the limit of $v/\tilde{v}\longrightarrow
0$.
As one can see, Eq.~(\ref{m}) does not give any further constraints to $C_1$.

Thus $C_1$ is  complex and unitary in the
limit of $v/\tilde v=0$ and  the quark mixing matrix $K$ is unitary in this
limit. Thus, at low energy, one unsuppressed Kobayashi--Maskawa CP phase is
introduced for the case of three generations.

\section{$\bar{\theta}$ calculation in a $Z_2$ model}
The one loop correction of $\bar{\theta}$ was calculated by Bento et.~al.
in the $Z_2$ Nelson--Barr type model \cite{bbp}. This calculation was
based on the work by Groffin et.~al. \cite{segre} who used the original formula
of Weinberg \cite{wei}. We briefly describe their calculation of the one loop
correction to $\bar{\theta}$ and compare their result with ours.
If we set the quark mass matrix as Eq.~(\ref{LR}) and set one loop
self energy as $\Sigma$, the one loop correction to $\bar{\theta}$ is
\begin{eqnarray}
\theta_{QFD}&=&\mbox{Arg Det}(M-\Sigma)_R\nonumber\\
&\simeq& - \mbox{Im tr}(M^{-1}\Sigma)_R.
\end{eqnarray}
The subscript $R$ means right--handed projection, $M_R=M_q$.
The general form of the self energy term which gives non--zero correction
to $\bar{\theta}$ is \cite{segre}
\begin{eqnarray}
\Sigma&=&-\frac{1}{16\pi^2}\sum_k\int^1_0 dx\left[-(1-x)M\gamma^0\Gamma_k
\gamma^0 + \Gamma_kM^\dagger\right]\nonumber\\
&&\times \ln[MM^\dagger x^2 + \mu_k^2(1-x)]\Gamma_k,
\end{eqnarray}
where $k$ runs over the mass eigenstates of scalars with mass $\mu_k$, and
$\Gamma_k$'s are Yukawa couplings matrices.
For the mass eigenbasis of scalars $(H_1, H_2, H_3)$, the mass matrix
$D=diag(\mu_1,\mu_2,\mu_3)$ is related with $M_{H}$ by $R$,
\begin{equation}
M_{H}=R^TDR. \label{D}
\end{equation}
This self energy term is equivalent to the Feynman diagram in Fig. 2.
Following Ref. \cite{bbp}, we decompose scalars as
\begin{eqnarray}
S&=&(V+s+it)e^{i\eta}/\sqrt{2}, \nonumber\\
H&=&(v+\rho+ia)/\sqrt{2},
\end{eqnarray}
where $V$ and $v$ are real and $a$ is absorbed to $Z^0$ boson, and $\eta$ is
the
relative phase of $S$ and $H$.
For the fermion doubling model, only the form of matrix $\Gamma$ and the form
of the quark mass matrix $M$ are different
from the Nelson--Barr type model.
\begin{eqnarray}
\Gamma_a\Phi^a&=&\left(
\begin{array}{cc}h_\rho&0\\ h_ss+ih_tt&0
\end{array}\right)\ \mbox{ Nelson--Barr}\nonumber\\
\Gamma_a\Phi^a &=&\left(
\begin{array}{cc}0&h_\rho\\ 0&h_ss+ih_tt
\end{array}\right)\ \mbox{ fermion doubling},
\end{eqnarray}
where $\Phi=(\rho,s,t)$, $h_s=be^{i\eta}+ce^{-i\eta}$ and
$h_t=be^{i\eta}-ce^{-i\eta}$, subscripts $U$ and $D$ in the real
coupling matrices $b$ and $c$ are omitted.
Note that $h$'s are $3\times 3$ matrices. Our previous $\beta$ and $\Omega$
in Sec. III are given by $\beta\equiv h_\rho v$ and $\Omega=h_s V$.
Diagonalization of the quark mass matrix $MM^\dagger$ by unitary
transformation,
we have \cite{bbp}
\begin{eqnarray}
\bar{\theta}&=& \frac{1}{16\pi^2}\sum_{a,b,k,A}\int^1_0 dx R^a_k R^b_k
\ln[M^{D2}_Ax^2+M^2_k(1-x)]\nonumber\\
&&\times{\rm Im}(U^\dagger_L\Gamma_aM_q^{-1}\Gamma_bM_q^\dagger U_L)_{AA}.
\label{theta}
\end{eqnarray}
One can verify that only the term containing $\Gamma_t M_q^{-1}
\Gamma_\rho$ contributes to the $\bar{\theta}$.
Bento et.~al. calculated the one loop correction \cite{bbp},
\begin{equation}
\bar{\theta}=\frac{1}{16\pi^2}(f^2-f^{\prime 2})(I_1+I_2+I_3),
\end{equation}
where $f$ and $f'$ are equivalent to $b$ and  $c$ of our model.
$I$'s are given in Ref. \cite{bbp}.
They showed that the integral $I$'s are proportional to $v/V$ if the
quartic couplings of $H-S$ have the same order of magnitude.
One can calculate the one loop correction in the fermion doubling model
by the same method.
To show it explicitly, we insert the following $M_q^{-1}$ and $U_L$ in
Eq.~(\ref{theta}):
\begin{equation}
M_q^{-1}=\left(\begin{array}{cc} -\alpha^{-1}\Omega\beta^{-1}& \alpha^{-1} \\
\beta^{-1}  &0\end{array}\right),\
U_L=\left( \begin{array}{cc} C_1&S_1\\ S_2&C_2 \end{array} \right).
\end{equation}
In Appendix A, we showed that $C_i^\dagger C_i\simeq 1$ for $i=1,2$,
from which we obtain
\begin{eqnarray}
\bar{\theta}&\simeq&\frac{1}{16\pi^2}R^\rho_k R^t_k\frac{1}{vV}
\mbox{tr}(C_2^\dagger h_t h_s^\dagger VC_2)
\times O(1) \nonumber\\
&\simeq&\frac{1}{16\pi^2}\mbox{tr}(b^2-c^2)R^\rho_k R^t_k\frac{V}{v}\times
O(1).
\end{eqnarray}
{}From Eq.~(\ref{pot}), we have scalar mass matrix for the basis $(h,s,t)$,
\begin{equation}
M_{H}\sim \left(\begin{array}{ccc} v^2& 2vV(\lambda_1+
2\lambda_2 \cos2\eta)& -2vV\lambda_2\sin2\eta\\ &
\sim V^2& \sim V^2\\ &&\sim V^2\end{array}\right)
\label{mh}
\end{equation}
If $\eta=0$, there is no CP violation in this model.
{}From Eq.~(\ref{mh}), it is easy to see that this is equivalent to
setting parameters as $\lambda_2\rightarrow0$ and $\lambda_1\rightarrow
\lambda_1+2\lambda_2$.
This shows that $\lambda_1$ is not relevant to $\bar{\theta}$ up to leading
order.
We approximate the rotation matrix $R$ which diagonalize $M_{H}$, for small
mixing between $H$ and $S$,
\begin{equation}
R\sim \left(\begin{array}{ccc}\sim 1& \epsilon& \epsilon'\\ \epsilon&
\cos\zeta& \sin\zeta\\ \epsilon'&-\sin\zeta&\cos\zeta\end{array}\right),
\end{equation}
where the $\epsilon$ and $\epsilon'$ are very small values due to the small
mixing angle.
One can easily verify from Eq.~(\ref{D}) that
\begin{equation}
\epsilon\sim\epsilon'\sim\lambda_2\frac{v}{V},
\end{equation}
where $R^\rho_kR^t_k\sim \epsilon$.
We get the same result of the $Z_2$ Nelson--Barr type model.
\begin{equation}
\bar{\theta}\sim\frac{1}{16\pi^2}(\tilde{b}^2-\tilde{c}^2)\lambda_2,
\end{equation}
where $\tilde{b}, \tilde{c}$ denote the orders of the
magnitudes.
In Sec. III, we discussed that $\lambda$'s have to be small $\sim v/V$ to avoid
cancellation problem in light Higgs scalar mass term.

\begin{figure}
\begin{picture}(350,220)(0,50)
\put (100,100){\line(1,0){280}}
\put (156,100){\circle*{2}}
\put (156.7,108.8){\circle*{2}}
\put (158.6,116.9){\circle*{2}}
\put (162,125.2){\circle*{2}}
\put (167,133.3){\circle*{2}}
\put (173,140.2){\circle*{2}}
\put (180,146.0){\circle*{2}}
\put (187,150.1){\circle*{2}}
\put (195,153.4){\circle*{2}}
\put (203,155.3){\circle*{2}}
\put (212,156){\circle*{2}}
\put (221,155.3){\circle*{2}}
\put (229,153.4){\circle*{2}}
\put (237,150.1){\circle*{2}}
\put (244,146.0){\circle*{2}}
\put (251,140.2){\circle*{2}}
\put (257,133.3){\circle*{2}}
\put (262,125.2){\circle*{2}}
\put (265.4,116.9){\circle*{2}}
\put (267.3,108.8){\circle*{2}}
\put (268,100){\circle*{2}}
\put (217.6,161.6){\circle*{2}}
\put (223.2,167.2){\circle*{2}}
\put (228.8,172.8){\circle*{2}}
\put (234.4,178.4){\circle*{2}}
\put (240,184){\circle*{2}}
\put (206.4,161.6){\circle*{2}}
\put (200.8,167.2){\circle*{2}}
\put (195.2,172.8){\circle*{2}}
\put (189.6,178.4){\circle*{2}}
\put (184,184){\circle*{2}}
\put (320.3,97.3){$\times$}
\put (179.5,181.3){$\otimes$}
\put (235.5,181.3){$\otimes$}
\put (170,185){$H$}
\put (245,185){$S,\sigma$}
\put (150,135){$H$}
\put (265,135){$S,\sigma$}
\put (125,85){$q_L$}
\put (208,85){$Q_R$}
\put (292,85){$Q_L$}
\put (349,85){$q_R$}
\end{picture}
\caption{The diagram which does not contribute to $\bar\theta$.}
\begin{picture}(250,300)(0,0)
\put (100,100){\line(1,0){280}}
\put (170,100){\circle*{2}}
\put (170.7,109.8){\circle*{2}}
\put (172.8,119.6){\circle*{2}}
\put (177,130.5){\circle*{2}}
\put (182.6,140){\circle*{2}}
\put (189.6,148.6){\circle*{2}}
\put (196.6,154.9){\circle*{2}}
\put (205,160.6){\circle*{2}}
\put (213.4,164.7){\circle*{2}}
\put (221.8,167.6){\circle*{2}}
\put (230.2,169.3){\circle*{2}}
\put (240,170){\circle*{2}}
\put (249.8,169.3){\circle*{2}}
\put (258.2,167.6){\circle*{2}}
\put (266.6,164.7){\circle*{2}}
\put (275,160.6){\circle*{2}}
\put (283.4,154.9){\circle*{2}}
\put (290.4,148.6){\circle*{2}}
\put (297.4,140){\circle*{2}}
\put (303,130.5){\circle*{2}}
\put (307.2,119.6){\circle*{2}}
\put (309.3,109.8){\circle*{2}}
\put (310,100){\circle*{2}}
\put (247,177){\circle*{2}}
\put (254,184){\circle*{2}}
\put (261,191){\circle*{2}}
\put (268,198){\circle*{2}}
\put (275,205){\circle*{2}}
\put (282,212){\circle*{2}}
\put (233,177){\circle*{2}}
\put (226,184){\circle*{2}}
\put (219,191){\circle*{2}}
\put (212,198){\circle*{2}}
\put (205,205){\circle*{2}}
\put (198,212){\circle*{2}}
\put (240,100){\circle*{2}}
\put (240,90){\circle*{2}}
\put (240,80){\circle*{2}}
\put (240,70){\circle*{2}}
\put (240,60){\circle*{2}}
\put (235.2,47){$\otimes$}
\put (132,85){$q_L$}
\put (201,85){$Q_R$}
\put (271,85){$Q_L$}
\put (342,85){$Q_R$}
\put (193.2,209.0){$\otimes$}
\put (277.3,209.0){$\otimes$}
\put (183,214){$H$}
\put (287,214){$S$}
\put (163,140){$H$}
\put (307,140){$S$}
\put (235,30){$S$}
\end{picture}
\caption{A self energy diagram which will give a non--zero contribution to
$\bar{\theta}$ in the $Z_2$ model.}
\end{figure}

\end{document}